\documentclass[preprint,3p,times]{elsarticle}

\usepackage{amssymb,amsmath,amsfonts,mathrsfs,bm,gensymb}
\usepackage{graphicx,float,subfig}
\usepackage{lineno}

\usepackage{geometry}
\usepackage{hyperref}

\usepackage{color}
\usepackage{soul}
\newcommand{\eqN}[1]{Eq.~(\ref{#1})}

\hypersetup{
    colorlinks=true,
    linkcolor=black,
    citecolor=black,
    filecolor=black,
    urlcolor=black,
}

\makeatother

\biboptions{sort&compress}

\begin{document}

\begin{frontmatter}

\title{Rayleigh wave propagation in nonlinear metasurfaces}

\author[add1]{A. Palermo\corref{corr1}}
\ead{antonio.palermo6@unibo.it}
\author[add2]{B. Yousefzadeh\corref{corr1}}
\ead{behrooz.yousefzadeh@concordia.ca}
\author[add3]{C. Daraio}
\author[add1]{A. Marzani}
\cortext[corr1]{Equal contributors and corresponding authors.}

\address[add1]{Department of Civil, Chemical, Environmental and Materials Engineering, University of Bologna, 40136 Bologna, Italy}
\address[add2]{Department of Mechanical, Industrial and Aerospace Engineering, Concordia University, Montreal, QC H3G 1M8, Canada.}
\address[add3]{Division of Engineering and Applied Science, California Institute of Technology, Pasadena, California 91125, USA.}

\begin{abstract}

We investigate the propagation of Rayleigh waves in a half-space coupled to a nonlinear metasurface. The metasurface consists of an array of nonlinear oscillators attached to the free surface of a homogeneous substrate. We describe, analytically and numerically, the effects of nonlinear interaction force and energy loss on the dispersion of Rayleigh waves. We develop closed-form expressions to predict the dispersive characteristics of nonlinear Rayleigh waves by adopting a leading-order effective medium description. 
In particular, we demonstrate how hardening nonlinearity reduces and eventually eliminates the linear filtering bandwidth of the metasurface. Softening nonlinearity, in contrast, induces lower and broader spectral gaps for weak to moderate strengths of nonlinearity, and narrows and eventually closes the gaps at high strengths of nonlinearity. We also observe the emergence of a spatial gap (in wavenumber) in the in-phase branch of the dispersion curves for softening nonlinearity. Finally, we investigate the interplay between nonlinearity and energy loss and discuss their combined effects on the dispersive properties of the metasurface. Our analytical results, supported by finite element simulations, demonstrate the mechanisms for achieving tunable dispersion characteristics in nonlinear metasurfaces. 

\end{abstract}


\end{frontmatter}
\section{Introduction}
\label{sec:intro}

The dynamics of elastic media equipped with a distribution of locally resonant elements, also known as metamaterials, has become an established field of research for mechanicians and physicists interested in manipulating the propagation of elastic waves. In the last two decades, a plethora of studies have proposed novel metamaterial designs to enrich the dynamics of discrete and continuous waveguides composed of spring-mass chains, rods,  beams,  plates, and 3D architectured lattices~\cite{hussein2014dynamics,ma2016acoustic}. 

For bulk metamaterials, the majority of the proposed layouts require complex geometries, thereby challenging the available manufacturing techniques both at the micro and at large scales. Therefore, more recent design strategies take advantage of \emph{elastic metasurfaces}, thin resonant interfaces patterned at the surface of an elastic waveguide or at the junction between different media~\cite{assouar}. Several of these resonant interfaces aim at manipulating the propagation of surface waves in elastic substrates, e.g., Rayleigh and Love waves. Metasurfaces of this type can support a rich variety of wave phenomena, such as amplitude attenuation via mode conversion (classical~\cite{colombi2016seismic} or Umklapp conversion~\cite{chaplain2020tailored}), energy trapping~\cite{dePonti2020}, waveguiding~\cite{addouche2014subwavelength}, and lensing~\cite{palermo2018control,fuentes2021design}. These phenomena are investigated to inform and enable technological applications based on surface acoustic waves (SAW) for signal processing~\cite{benchabane2019elastic},  energy harvesting~\cite{dePonti2020}, and mitigation devices for ground-borne vibrations~\cite{palermo2016engineered}. 

Theoretical models of increasing complexity have been developed to describe the dynamics of elastic metasurfaces. At its simplest level, a metasurface can be modelled as a uniform array of discrete oscillators (a spring-mass system) attached to a homogeneous isotropic elastic substrate~\cite{Garova}. The uniform distribution and the sub-wavelenght dimensions of the resonators allow a mathematical formulation based on the effective-medium theory, where the resonators are included as dynamic effective boundary conditions of the Robin type for the substrate~\cite{Garova,maznev2015waveguiding}. This formulation can describe the dynamic coupling between the resonators and the substrate. The dynamics of the resonators has later been enriched by adopting continuous models for rods~\cite{colquitt2017seismic}, beams~\cite{wootton2019asymptotic} and plates~\cite{marigo2020effective}. Similarly, the description of the substrate has been improved to account for more complex rheological models such as the presence of fluids or the variation of the elastic properties along the medium depth~\cite{pu2020seismic}. 

Alternatively, asymptotic analysis can be used to obtain closed-form expressions for dispersion relations, for example for longitudinal and flexural metasurfaces ~\cite{colquitt2017seismic,wootton2019asymptotic}. Similar techniques can be utilized to obtain homogenized models for calculation of the wave fields and reflection coefficients of periodic metasurfaces~\cite{marigo2020effective}. Additionally, asymptotic analysis can be successfully used to describe the nonreciprocal propagation of Rayleigh waves in metasurfaces with space-time modulated properties~\cite{wu2021non,palermo2020surface}.

Graded arrangements of resonators have been investigated  for wave guiding and focusing applications~\cite{dePonti2020,metawedge}. When the arrangement of surface resonators is not regular in space, Green's functions and multiple scattering techniques can be used to calculate the wave field and the transmission coefficients~\cite{pu2021lamb}. 

Despite the rich literature on elastic metasurfaces, the focus has predominantly remained on the propagation of low-amplitude  surface waves across linear locally resonant structures. Indeed, nonlinearity may be utilized to introduce new functionalities for metasurfaces. Nonlinear effects can be triggered  by generation of finite-amplitude waves or by engineering the interaction force between the resonator and substrate~\cite{TournatBertoldi,PalermoCelli}. The presence of nonlinear forces can enable and enhance a variety of unique functionalities such as tunable~\cite{Pai} and asymmetric~\cite{NRM} propagation of waves. Although nonlinearity has been utilized to expand the available design space in metamaterials~\cite{hussein2014dynamics,KochmannBertoldi}, its application in metasurfaces remains rare~\cite{assouar,bahram,mu2020review}. The most relevant studies, to the best of our knowledge, are those that involve nonlinear resonators attached to a one-dimensional substrate ~\cite{WallenNL,casalotti2018metamaterial,beamHarmonicGeneration,lou2020revealing,onemore}.

In this work, we investigate the nonlinear dispersive properties of a metasurface consisting of a uniform array of local resonators placed on the free surface of a homogeneous elastic substrate. Both nonlinearity and energy loss of the local resonators are considered. We focus on the effects of the nonlinear interaction between the surface waves and resonators, and describe how the amplitude-dependent response of the surface resonators influence the dispersion of Rayleigh waves. To this purpose, we develop closed-form expressions for the leading-order nonlinear dispersion relation of the metasurface and discuss the effects of different types of nonlinearity (softening and hardening), of energy loss, and of the combined effects of damping and nonlinearity. We support and validate our theoretical findings by reproducing the main dispersion characteristics using direct numerical simulations. The numerical results highlight additional nonlinear effects (third harmonic generations) that are not accounted for in our analytical model. 

This work is organized as follows:  the general formulation of the problem is provided in Sec.~\ref{sec:formulation}. In Sec.~\ref{sec:alldispersion}, we discuss the influence of the surface resonators on the dispersion of Rayleigh waves and highlight the prominent effects of nonlinearity in the interaction force. The influence of incorporating energy loss in the surface resonators is explored in Sec.~\ref{sec:damping} for both linear and nonlinear resonators. Numerical validation of the main results are presented in Sec.~\ref{sec:numerics}. We summarize our findings in Sec.~\ref{sec:conclusion} and discuss a few avenues for further developments.

\section{General formulation}
\label{sec:formulation}

\subsection{Metasurface model}
\label{sec:model}

\begin{figure*}
    \includegraphics[trim={0 2cm 0 2cm}, clip,width=\linewidth]{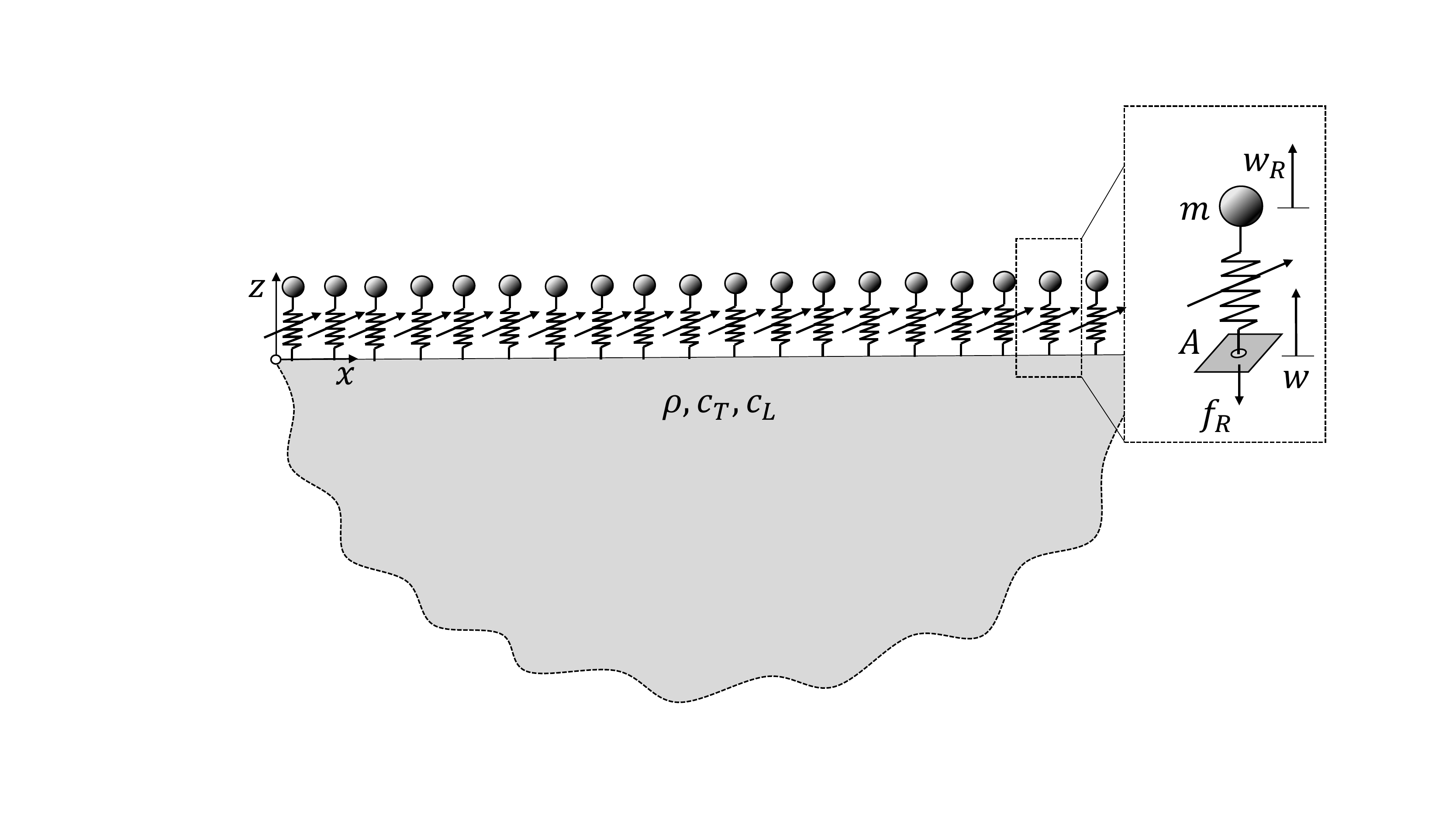}
\caption{Schematic of the setup.}
\label{f:schematic}       
\end{figure*}

The system comprises a uniform array of nonlinear resonators placed on a semi-infinite, isotropic elastic substrate (see Fig.~\ref{f:schematic}). In a 2D plane-strain state, the elastic waves in the substrate are governed by the equations 
\begin{equation}
	\label{nabla2}
	\begin{aligned}
		\phi_{,xx}+\phi_{,zz} &= \phi_{,tt}/c_L^2 \\
		\psi_{,xx}+\psi_{,zz} &= \psi_{,tt}/c_T^2
	\end{aligned}
\end{equation}
 where $\phi$ and $\psi$ are, respectively, the potentials for the longitudinal and shear waves,  $c_L=\sqrt{(\lambda+2\mu)/\rho}$ and $c_T=\sqrt{\mu/\rho}$ are the corresponding wave speeds~\cite{graff,ewing}, with $\rho$, $\lambda$  and  $\mu$, respectively, the density and the Lam{\'e} parameters of the substrate. We denote with $(.)_{,j}$  the partial derivative  $\partial(.)/\partial j$.
 
 For a substrate with a surface at $z=0$ and extending to $z=-\infty$, we assume harmonic wave solutions to \eqN{nabla2}  that travel in the $x$ direction and remain bounded in the $z$ direction
\begin{subequations}
	\label{phipsi}
	\begin{align}
		\label{phi}
		\phi(x,z,t) &= B_L e^{q_L z} e^{i(\omega t-qx)} \, , \, q_L = \sqrt{q^2-\omega^2/c_L^2} \\
		\label{psi}
		\psi(x,z,t) &= B_T e^{q_T z} e^{i(\omega t-qx)}\, , \, q_T = \sqrt{q^2-\omega^2/c_T^2}
	\end{align}
\end{subequations}
where $B_L$ and $B_T$ are the potential amplitudes. The stress components in the substrate are related to the potentials $\phi$ and $\psi$ by
\begin{subequations}
	\begin{align}
		\label{sigzx}
			\sigma_{zx} &= \mu (2\phi_{,xz} + \psi_{,xx} - \psi_{,zz}) 
		\\
		\label{sigzz}
			\sigma_{zz} &= (\lambda+2\mu) (\phi_{,zz} + \psi_{,xz}) + \lambda (\phi_{,xx} - \psi_{,xz}) 
	\end{align}
\end{subequations}
The relation between the wavenumber $q$ and frequency $\omega$ of surface waves, i.e., the dispersion relation, is determined from the boundary conditions. In this regards, we assume the resonators  far enough from each other that there is no contact between them, and consider only their vertical motion. Then, the following equations  describe the boundary conditions for the substrate free surface ($z=0$):
\begin{subequations}
	\label{eq:BD1}
	\begin{align}
		\label{BDzx}
		&\sigma_{zx}(x,0,t) = 0 \\
		\label{BDzz}
		&\sigma_{zz}(x,0,t)A = f_R(y) 
	\end{align}
\end{subequations}
where  $f_R(y)$ is the force exerted by the resonator of mass $m$ on the substrate and distributed uniformly over the effective contact area $A$. In \eqN{BDzz}, it is implicitly assumed that the distance between adjacent resonators is smaller than the wavelength $2\pi/q$ of the surface waves so that the resonators can be treated as an effective continuum~\cite{Boechler}. The interaction force $f_R(y)$ depends on the relative motion between the resonator mass $w_R(x,t)$ and the vertical motion of the substrate surface $w(x,0,t)$:
\begin{equation}
\label{yVw}
	y(x,t)=w_R(x,t)-w(x,0,t)=\left(\frac{1}{r}-1\right) w(x,0,t)
\end{equation}
where $r=\frac{w(x,0,t)}{w_R(x,t)}$.
Thus, the equilibrium equation of the local resonator reads
\begin{equation}
	 \label{BDres}
        m\ddot{y}+f_R(y)=-m\ddot{w}(x,0,t)\\
\end{equation}

The four equations (\ref{BDzx}), (\ref{BDzz}),  (\ref{yVw}) and (\ref{BDres}) are used to determine the dispersion relation for surface waves defined in~\eqN{phipsi}. When the resonator is linear, it is not necessary to introduce the relative motion $y(x,t)$ because $r$ does not depend on $w(x,0,t)$. In the nonlinear problem, instead, the parameter $r$ is amplitude-dependent, e.g.  $r=r(y(w))$. Thus, the use of relative motion is necessary (see Sec.~\ref{sec:alldispersion}).

\subsection{Interaction force and resonator response}
\label{sec:cubic_sol}

For finite, small-amplitude oscillations of the resonators on the substrate (weak nonlinearity), the interaction force may be approximated by a Taylor expansion around the equilibrium position of the resonators,
\begin{equation}
    \label{taylor}
    f_R \approx k_0 + k_1 y + k_2 y^2 + k_3 y^3
\end{equation}
The values of the stiffness coefficients in \eqN{taylor} are determined by the specific contact law, such as the Hertz law for contact bead resonators~\cite{Boechler}. Here, we consider the influence of these terms individually, without tying $f_R$ to a specific interaction force. This general discussion can be narrowed down to a particular nonlinear interaction force if necessary, as long as using a Taylor expansion is a reasonable approximation for that force.

The static term, $k_0$, shifts the equilibrium position of the resonators. One can think of this as caused by a precompression of the resonator onto the substrate, which can be utilized in tuning the interaction force~\cite{PalermoCelli}. When the  force $f_R(y)$ is not subjected to tuning, without loss of generality, one can choose the origin of $y$ such that $k_0=0$. We take this approach throughout this work. The linear stiffness, $k_1$, is the leading-order term. Depending on the properties of the interaction force, the next term could be either the quadratic stiffness, $k_2$, or the cubic stiffness, $k_3$. 

As a first step, we consider an interaction force for which the quadratic term is negligible with respect to the cubic term, $k_2/k_3 \rightarrow 0$. 
{\color{black} Ignoring the quadratic term simplifies the analysis and enables us to focus on the most typical effects of nonlinear interaction force first. We discuss the prominent effects of quadratic nonlinearity in Sec.~\ref{sec:alldispersion}. }

The interaction between the nonlinear resonator and surfaces waves is described by~\eqN{BDres}.
The vertical displacement of the substrate satisfies the relation 
\begin{equation}
	\label{wphipsi}
	w(x,z,t) = \phi_{,z}+\psi_{,x}
\end{equation}
Using \eqN{phipsi}, we have $w(x,0,t)=B_W\exp(i(\omega t - qx))$ for the substrate, with
\begin{equation}
	\label{BW}
	B_W=q_LB_L-iqB_T
\end{equation}
The term on the right-hand side of~\eqN{BDres} can be treated as a harmonic excitation. Accordingly, \eqN{BDres} can be viewed as a harmonically forced Duffing oscillator, the solutions of which are well documented~\cite{NayfehMook}. In this work, we assume a harmonic wave solution for the relative motion of the resonators, $y(x,t)=Y\exp(i(\omega t - qx))$, consistent with \eqN{phipsi} and \eqN{yVw}. Thus, we can relate $Y$ and $B_W$ in form of an algebraic equation
\begin{equation}
    \label{YBW}
    \left(\Omega-\omega^2\right)Y=\omega^2B_W
\end{equation}
where
\begin{equation}
    \label{Omega}
    \Omega=\omega_R^2(1+3\beta|Y|^2)
\end{equation}
and  $\beta=k_3/k_1$. As expected, we retrieve the linear response when $\beta=0$. The interaction force can then be approximated as
\begin{equation}
    \label{fRcubic}
    f_R = m\Omega Y\exp(i(\omega t - qx))
\end{equation}

In writing \eqN{Omega} and \eqN{fRcubic}, we have implicitly restricted our analysis to the motion of the resonators near their primary resonance. This assumption is in accordance with a single-term plane-wave expansion of the wave field. Note that the adopted leading-order approximation ignores the possibility of the generation of higher (or lower) harmonics due to nonlinearity, and subsequently ignores any modal interaction with higher-order plane waves. The most accessible higher-order effect for cubic nonlinearity is the generation of third harmonics, which we briefly discuss in Sec.~\ref{sec:numerics}. See~\cite{beamHarmonicGeneration} for a detailed example of third-harmonic generation in a similar problem in which the substrate is a thin beam. 

\section{Dispersion of Rayleigh waves}
\label{sec:alldispersion}
To obtain the dispersion relation in the presence of weak cubic nonlinearity, we follow the classical approach developed for linear metasurfaces. In a linear system, the four equations (\ref{BDzx}), (\ref{BDzz}), (\ref{yVw}) and (\ref{BDres}) yield a homogeneous algebraic system of equations for the four wave amplitudes $B_L$, $B_T$, $B_W$ and $Y$. A non-trivial solution is possible if the following relation holds
\begin{equation}
    \label{dispersion}
    r \left(4q^2q_Lq_T-(q^2+q_T^2)^2\right) \mu^2 A = m \rho q_L\omega^4
\end{equation}
where 
\begin{equation}
    \label{ratio}
    r=\frac{w(x,0,t)}{w_R(x,t)}=1-\frac{\omega^2}{\Omega}
\end{equation}
and $\Omega=\omega_R^2=k_1/m$ in the linear, undamped problem. When the interaction force is linear, \eqN{dispersion} corresponds to the classical dispersion relation reported by Boechler et al.~\cite{Boechler}. 
\eqN{dispersion} can be easily adapted to account for the nonlinear interaction force: the amplitude-dependent dynamics of the surface resonators are introduced through parameter $r$, defined in \eqN{ratio}. For cubic nonlinearity, the amplitude-dependent behavior is captured by $\Omega=\Omega(\beta,|Y|)$ in \eqN{Omega}. Naturally, the nonlinear effects depend on both the amplitude of motion (e.g., quantified by $Y$ or $B_W$) and the value of the cubic stiffness $\beta$. 

In this work, we fix $\beta=\pm 1$ and tune the amplitude-dependent dynamics by imposing $B_W$. This choice is  relevant to experimental realizations of the setup, where $B_W$ may be controlled directly as the input to the system~\cite{PalermoCelli}. Qualitatively, the same behavior may be observed by fixing the wave amplitude and varying $\beta$ instead. 

\begin{figure}[H]
    \includegraphics[width=\linewidth]{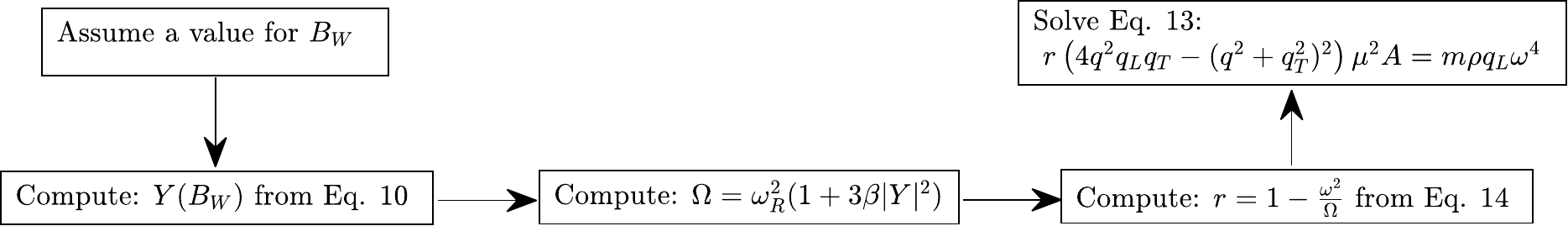}
\caption{Flow chart depicting the procedure to calculate the amplitude-dependent dispersive properties of the nonlinear metasurface.
}
\label{f:flowchart}       
\end{figure}

Hence, the leading-order effect of the cubic stiffness can be captured in our analysis by changing $\Omega$ from $\omega_R^2$ to $\omega_R^2(1+3\beta|Y|^2$) in Eqs.~(\ref{ratio}). To obtain the dispersion relation for a given amplitude of the incoming wave $B_W$, we calculate the relative motion $Y$ using Eqs.~(\ref{YBW}) and (\ref{Omega}). The amplitude-dependent dispersive properties are then obtained by solving the dispersion relation in \eqN{dispersion}. The flowchart in Fig.~\ref{f:flowchart} summarizes this procedure.

{\color{black}Concerning the quadratic nonlinearity,  one can still use the proposed procedure  to obtain an amplitude-dependent dispersion relation when the quadratic term in \eqN{taylor} is not negligible with respect to the cubic term. In this case,  a standard multiple-scale analysis can be employed to solve \eqN{BDres} for the relative motion~\cite{NayfehMook}. The leading-order effect would appear then as a correction to the parameter $\beta$ in \eqN{Omega}. This analysis would be based on the assumption of weak nonlinearity (finite, small-amplitude waves), and would ignore the modal interaction between plane waves. The typical higher-order effects for quadratic nonlinearity are the generation of second harmonics and the appearance of a DC shift in the oscillations (also known as drift). We postpone a detailed investigation of these effects to future studies. We discuss the details of the dispersion curves for both softening and hardening nonlinearities in Sec.~\ref{sec:cubic_dispersion}}

\subsection{Linear analysis} 
\label{sec:linear}

\begin{figure}
    \includegraphics[width=\linewidth]{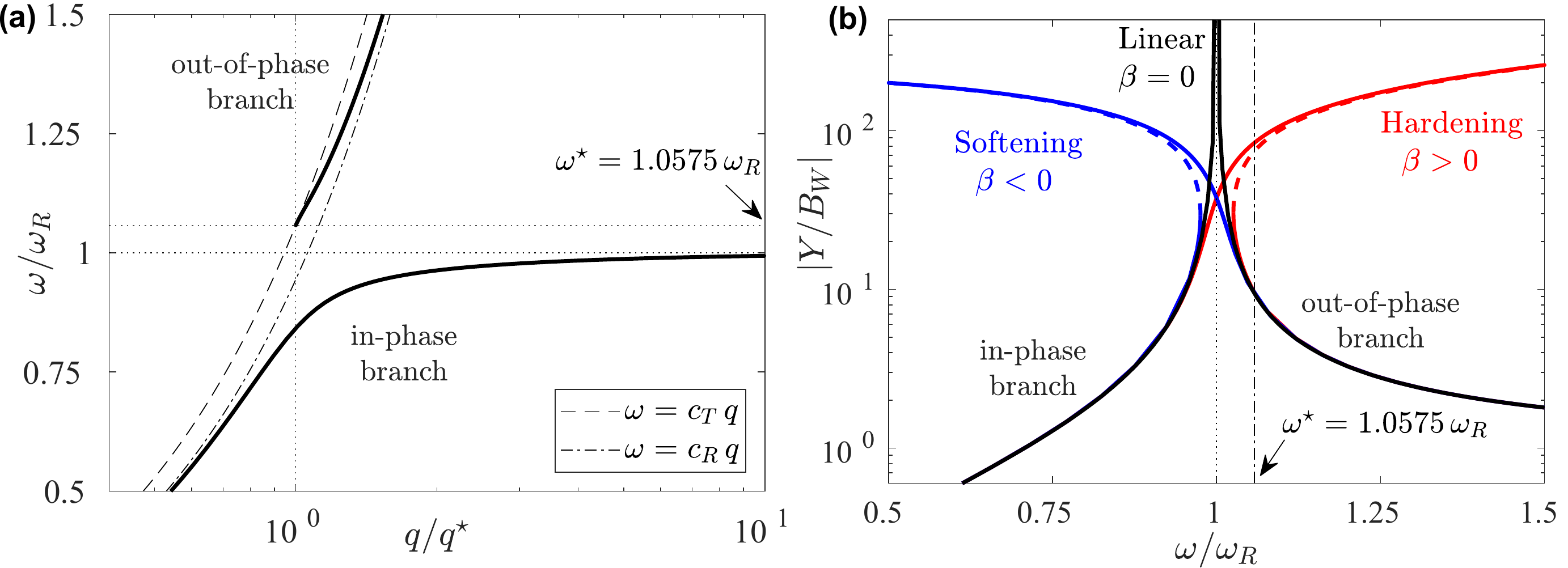}
\caption{Panel (a): dispersion relation for Rayleigh waves with linear interaction force, $c_L/c_T=1.5$ and $m\omega_R/\rho A c_T=0.15$. The dash-dotted curve corresponds to $\omega = c_R q$, the dispersion curve for Rayleigh waves in the absence of surface resonators. The dashed curve corresponds to shear wave dispersion, $\omega = c_T q$, above which waves are not bound to the surface. The horizontal dotted lines indicate the lower bound ($\omega=\omega_R$) and the upper bound ($\omega=\omega^\star$) of the band gap. The vertical dotted line indicates $q=q^\star$, the onset of the out-of-phase branch of the dispersion curve. Panel (b): steady-state dynamics of an undamped surface resonator, $|Y/B_W|$, as a function of the frequency of the incoming Rayleigh wave, $\omega$, with linear ($\beta=0$) and nonlinear ($\beta=\pm 1$) interaction forces. The incoming wave amplitude is $B_W=0.005$ for both the hardening and softening types of nonlinearity. The unstable portions of the response curves are depicted using dashed lines. The vertical dash-dotted line indicates $\omega=\omega^\star$.
}
\label{f:disp_linear}       
\end{figure}

To better appreciate the effect of nonlinerity on the metasurface dispersion curve, we briefly summarize the main dispersive features of a linear metasurface; i.e. $f_R(y) = k_1 y$. The presence of the local resonators creates a band gap in the dispersion curve, $\omega_l < \omega_{gap} < \omega_u$, for the propagation of surface waves (see Fig.~\ref{f:disp_linear}a). The lower bound $\omega_l=\sqrt{\Omega}$ coincides with the horizontal tangency of the dispersion curve at $(q,\omega)=(+\infty,\omega_R)$. This corresponds to the resonance of the surface oscillator subject to harmonic base displacement, as depicted in Fig.~\ref{f:disp_linear}b for $\beta=0$.
We can see from \eqN{ratio} that for frequencies below the local resonance, $\omega<\omega_R$, the resonators move in phase with the substrate ($r>0$). The upper bound, $\omega_u$, satisfies $\omega=qc_T$, and is found to be the positive root of
\begin{equation}
    \label{wupper}
    \omega_u^2-\frac{m}{\rho A} \left(c_T^{-2}-c_L^{-2}\right)\Omega\omega_u-\Omega=0
    \,,\quad 
\end{equation}
According to \eqN{phipsi}, the solutions of \eqN{dispersion} for which $\omega>qc_T$ correspond to waves that are not confined to the surface of the substrate. For all the admissible surface waves above the band gap, the resonators move out of phase with the substrate ($r<0$). 

We choose $c_L/c_T=1.5$ and $\frac{m\omega_R}{\rho A c_T}=0.15$ throughout this work. These are the two key parameter groups that determine the physics of linear wave propagation. The first group determines the elastic properties of the substrate and the second one determines the strength of the interaction between the resonators  the substrate. The upper bound of the band gap, determined from \eqN{wupper} to be at frequency $\omega^\star=1.0575\omega_R$, with a corresponding wavenumber $q^\star=c_T\omega^\star$, are used to normalize the dispersion curves throughout this work (see Fig.~\ref{f:disp_linear}a). The reader can refer to~\cite{Boechler,PalermoSeb,WallenMaznev} for further details on the physics of the linear problem.

\subsection{Nonlinear analysis}

\label{sec:cubic_dispersion}

\subsubsection{Hardening nonlinearity}
\label{sec:cubic_hard}

\begin{figure}
    \includegraphics[width=\linewidth]{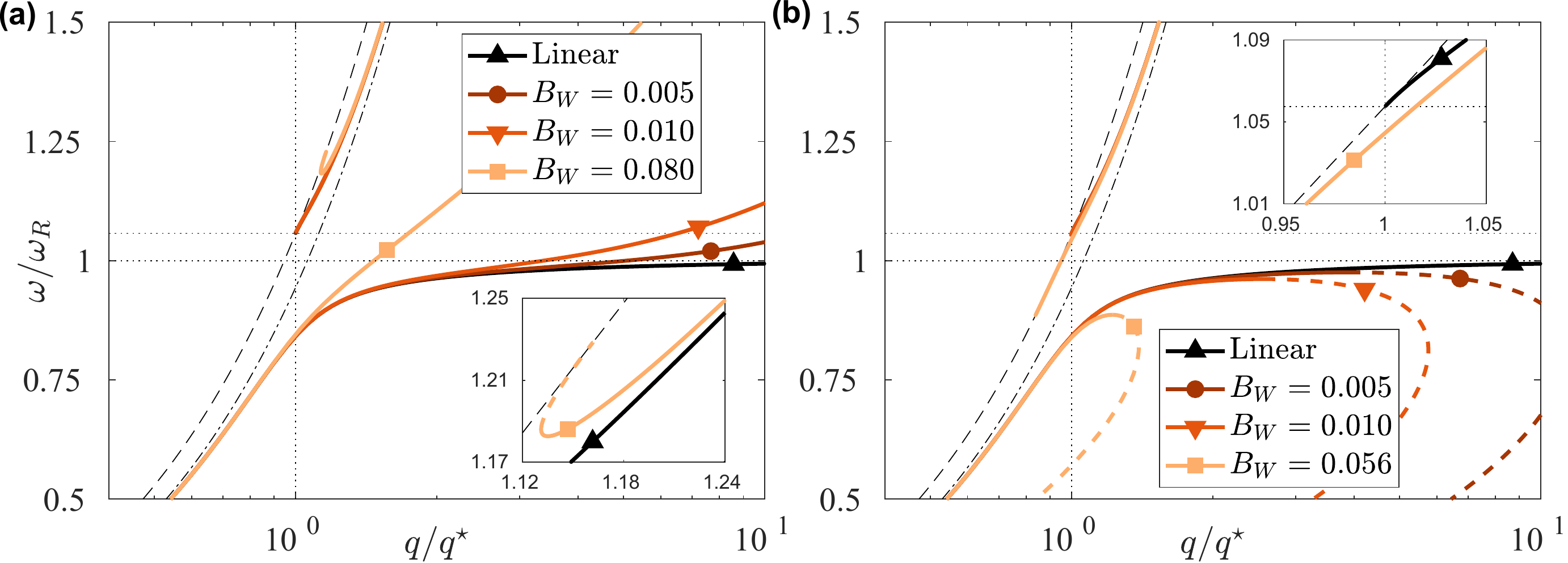}
\caption{Panel (a): influence of hardening, cubic nonlinearity ($\beta=1$, $k_2=0$) on the dispersion of Rayleigh waves for different values of the incoming wave amplitude, $B_W$. The onset of the out-of-phase branch remains almost unchanged for small values of $B_W$. The inset highlights the effect of nonlinearity on the onset of the out-of-phase branch at higher values of $B_W$.  Panel (b): influence of softening, cubic nonlinearity ($\beta=-1$, $k_2=0$) on the dispersion of Rayleigh waves for different values of the incoming wave amplitude, $B_W$.  The onset of the out-of-phase branch remains almost unchanged for small values of $B_W$. The inset highlights the effect of nonlinearity on onset of the out-of-phase branch at higher values of $B_W$. In both the panels, the linear dispersion curve from Fig.~\ref{f:disp_linear} is included for comparison and the unstable portions of the response curves are depicted using dashed lines.
}
\label{f:disp_hard}       
\end{figure}

Fig.~\ref{f:disp_hard}a shows the influence of hardening cubic nonlinearity ($\beta=1$) on the dispersion curves. Focusing on the in-phase (acoustic) branch, one can note that $\omega=\omega_R$ is no longer the horizontal asymptote of the dispersion curve. The frequency of the in-phase branch increases monotonically with its wavenumber and the band gap disappears. This behavior is driven by the response of the surface resonator to a prescribed base motion (Fig.~\ref{f:disp_linear}b), which presents the well-known bending at its primary resonance. Because there is no damping, the in-phase branch in Fig.~\ref{f:disp_linear}b (extending to the origin) increases monotonically with frequency, dictating a similar behavior in the corresponding in-phase branch of the dispersion curve.

Conversely, the onset of the out-of-phase (optical) branch remains almost unchanged for incoming base motion with very low amplitude (Fig.~\ref{f:disp_hard}, $B_W=0.005$ and $B_W=0.010$). This is linked to the behavior of the out-of-phase branch of the resonator's response curve in Fig.~\ref{f:disp_linear}b, i.e., the branch that becomes asymptotically horizontal ($|Y/B_W|\rightarrow 1$ when $\omega\rightarrow \infty$). 
Since waves for which $\omega>qc_T$ are not confined to the surface, the out-of-phase branch  terminates at $\omega=\omega^\star$, as shown in Fig.~\ref{f:disp_hard}a. 
For larger values of $B_W$, the turning point in the out-of-phase branch occurs at a frequency higher than $\omega^\star$. A corresponding turning point appears in the dispersion curve, as shown in the inset of Fig.~\ref{f:disp_hard}a for $B_W=0.080$. This turning point separates stable Rayleigh waves from unstable ones. 

In summary, the linear band gap between $\omega_R<\omega<\omega^\star$ disappears in the presence of hardening nonlinearity. The in-phase branch of dispersion is much more sensitive to nonlinearity (controlled by the amplitude of incoming wave, $B_W$) than the out-of-phase one. We anticipate that the presence of damping can have a significant influence on the dynamics near the resonance, as will be discussed in Sec.~\ref{sec:damping}. 

\subsubsection{Softening nonlinearity}
\label{sec:cubic_soft}

The influence of softening cubic nonlinearity  ($\beta=-1$) on the dispersion curve is shown in Fig.~\ref{f:disp_hard}b. We observe that the in-phase branch turns back on itself, consistently with the behavior of the response curve in Fig.~\ref{f:disp_linear}b. The group velocity is zero at the turning point (i.e., $\partial \omega / \partial q=0$), beyond which the solution is unstable. Therefore, the in-phase branch of dispersion terminates at the turning point and the lower bound of the band gap effectively decreases as a function of the input wave amplitude. Note in Fig.~\ref{f:disp_hard}b that the change in admissible wavenumbers for the in-phase branch is much greater than the change in admissible frequencies. In other words, the lower bound of the band gap occurs at a finite wavenumber (in contrast to a linear system). This results in the creation of a partial gap in admissible wavenumbers for surface waves moving in phase with the substrate. 

Similar to what we observed for the hardening metasurface, the onset of the out-of-phase branch of dispersion remains almost unchanged when the incoming waves have very low amplitude (Fig.~\ref{f:disp_hard}b, $B_W=0.005$ and $B_W=0.010$). 
This is because the dynamics of the surface resonator has an insignificant dependence on $B_W$ at frequencies near and above $\omega^\star$ (see Fig.~\ref{f:disp_linear}b). In this range of parameters, the effective band gap of the system becomes wider as $B_W$ increases. For larger values of $B_W$, the onset of the out-of-phase dispersion branch occurs at increasingly lower frequencies, as expected from the softening nature of nonlinearity. The band gap becomes increasingly narrower, as a result. The band gap is eventually closed if the amplitude of the incoming wave is large enough, as depicted in Fig.\ref{f:disp_hard}b for $B_W=0.056$. 

In summary, the linear band gap between $\omega_R<\omega<\omega^\star$ first widens and then disappears as a function of the incoming wave amplitude. When a band gap exists, its lower bound occurs at a finite wavenumber.

\section{Energy loss in resonators}
\label{sec:damping}

Very low amount of damping does not significantly alter or interfere with the interaction between surface waves and resonators in the linear operating range, as evidenced by the success of previous experimental studies~\cite{Boechler,PalermoSeb}. Energy loss is nevertheless inevitable in any experimental realization of the setup in Fig.~\ref{f:schematic}. In this section, we investigate the effect of energy loss for both linear and nonlinear resonators. 

We focus solely on energy loss that occurs in the oscillators (i.e., the interaction force), which is modeled as a linear, viscous damping mechanism acting in parallel to the elasticity. This is incorporated in the interaction force by adding the term $c\dot{y}$ to \eqN{BDres} and \eqN{taylor}, where $c$ is the damping coefficient. 
This additional term appears in the dispersion relation by adding $ic\omega$ to parameter $\Omega$ in \eqN{ratio}. The effects of incorporating energy loss in the substrate are discussed elsewhere for linear waves~\cite{Djafar2021}. 

In our analysis of harmonic wave propagation in a damped periodic medium, we adopt the convention of using real-valued wave frequency and allowing the wavenumber to become complex-valued~\cite{GeneralMead,mikeJAP,leamy,laudePRB1}. This formulation is particularly suitable when analyzing the effects of damping due to continuous harmonic wave excitation, i.e., a wave transmission problem. The alternative formulation with complex-valued frequency and real-valued wavenumber would give the same results at long wavelengths and for small values of damping~\cite{jensenJVA}. Physically, however, the formulation with real-valued wavenumbers corresponds to a different experimental scenario: a wave scattering experiment in which a prescribed wavenumber is given as the input to the system~\cite{Maznev}. Although both approaches are experimentally viable~\cite{Boechler,PalermoSeb}, we base our analysis on the transmission problem due to its relative ease of implementation in experiments. Further distinctions between the two approaches are discussed in Ref. ~\cite{Maznev}.

\begin{figure*}
    \includegraphics[width=\linewidth]{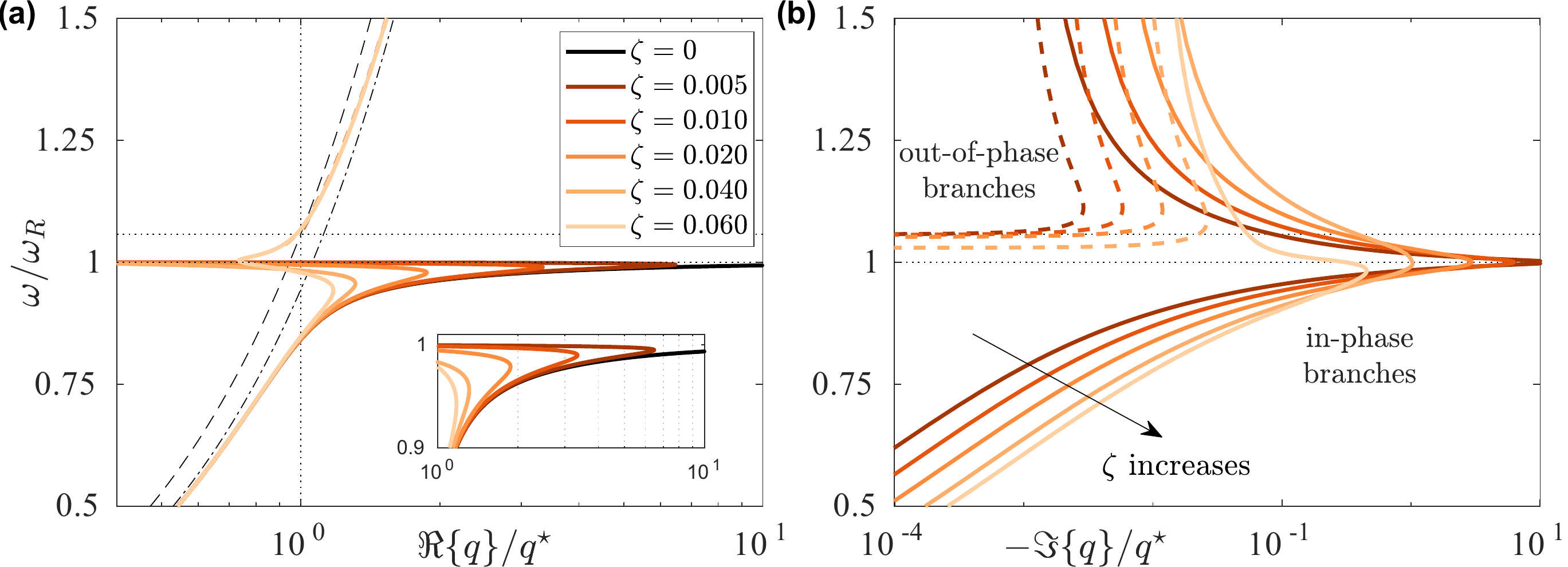}
\caption{Influence of energy loss on the linear dispersion of Rayleigh waves for increasing values of damping ratio. Panels (a) and (b) show, respectively, the real and imaginary parts of the dispersion curves. The imaginary part is shown with a negative sign. Notice that the dispersion curve of the undamped system ($\zeta=0$) is real-valued. 
}
\label{f:disp_zett}       
\end{figure*}

\subsection{Linear damped metasurface}
\label{sec:damping_linear}

Energy loss appears in the dispersion relation through the amplitude ratio $r$, defined in \eqN{ratio}. In the linear operating range, we have $\Omega=\omega_R^2(1+2i\zeta\omega/\omega_R)$, where $\zeta=c/2\omega_R$ is the damping ratio. The damping force introduces a gradual, frequency-dependent phase difference between the motions of the resonator and substrate under it. This is captured in our formulation by complex-valued wave amplitudes. More importantly, the wavenumber associated with a given incoming wave becomes complex-valued as well. The imaginary part of the wavenumber represents spatial decay (due to damping) of the wave amplitude as the wave propagates along the surface. 

Fig.~\ref{f:disp_zett} shows the evolution of the surface wave dispersion curves for increasing values of damping ratio; panel~(a) shows the real part of the wavenumber and panel~(b) shows its imaginary part with a negative sign. As expected, introduction of energy loss in the resonator response results in spatial decay of the surface waves. In particular, the real part of the in-phase branch bends back on itself at a finite value of the wavenumber; see Fig.~\ref{f:disp_zett}(a) and its inset. This creates an upper limit for the wavenumber of surface waves moving in phase with the local resonators. The emergence of partial wavenumber gaps due to damping is well documented in discrete and continuous models of phononic crystals~\cite{mikeJAP,leamy}. The onset of the partial wavenumber gap corresponds to the maximum displacement amplitude of the local resonators, and occurs at longer wavelengths as damping increases. Moreover, increasing the value of damping  slightly reduces the frequency at which the in-phase branch folds on itself, consistent with the viscous nature of the damping force.

The influence of damping on the out-of-phase branch is better observed in the imaginary part of the dispersion curve, Fig.~\ref{f:disp_zett}(b). When damping is small, the magnitude of the imaginary parts of the out-of-phase branches are smaller than those of the in-phase branches, meaning that the spatial decay rate is smaller for the out-of-phase branch. This is no longer the case for moderate values of damping: the two branches merge together near $\zeta=0.060$ (quality factor of 8.3). Similar effects are reported in previous studies~\cite{Maznev} and observed in experiments with sound waves~\cite{FangDamped}. 
In the presence of damping, the upper edge of the band gap (denoted by $\omega^\star$ when $\zeta=0$) occurs at a frequency lower than $\omega^\star$. This can be observed in Fig.~\ref{f:disp_zett}(b) by noting the horizontal asymptotes of the out-of-phase branches. This effect could contribute to a slight widening of the band gap.

\subsection{Nonlinear damped metasurface}
\label{sec:damping_nonlinear}

Following the approaches in Sec.~\ref{sec:cubic_sol} and Sec.~\ref{sec:damping_linear}, we now incorporate the effects of energy loss and nonlinearity in the dispersion relation by using  $\Omega=\omega_R^2(1+2i\zeta\omega/\omega_R+3\beta|Y|^2)$ in \eqN{YBW} and \eqN{ratio}. The rest of the analysis follows the same procedure as in Sec.~\ref{sec:cubic_dispersion}. 

\begin{figure*}[h!]
    \includegraphics[width=\linewidth]{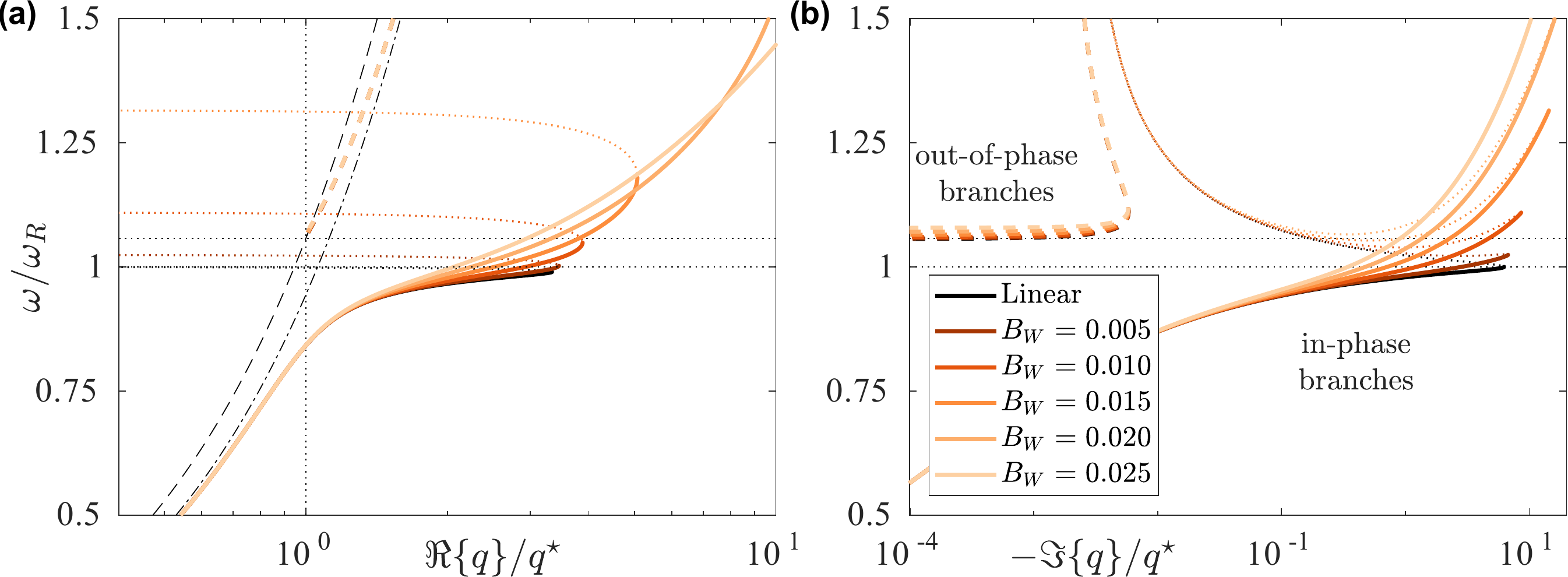}
\caption{Influence of hardening, cubic nonlinearity ($\beta=1$, $k_2=0$) on the dispersion of damped Rayleigh waves with $\zeta=0.010$ for different values of the incoming wave amplitude, $B_W$. Panels (a) and (b) show, respectively, the real and imaginary parts of the dispersion curves. The linear dispersion curve from Fig.~\ref{f:disp_zett} is included for comparison. The portions of the in-phase branches beyond the first turning point are depicted in dotted lines {\color{black}to highlight the maximum admissible wavenumber in each case}. 
}
\label{f:disp_zett_hard}       
\end{figure*}

Fig.~\ref{f:disp_zett_hard} shows the influence of hardening cubic nonlinearity on the dispersion of damped surface waves with $\zeta=0.010$. In contrast to the undamped dispersion curves (cf. Fig.~\ref{f:disp_hard}), all the in-phase branches fold back on themselves in the presence of damping -- recall the discussion in Sec.~~\ref{sec:damping_linear}. Due to the hardening nature of the nonlinear interaction force ($\beta>0$), the locus of the fold point moves to higher frequencies and shorter wavenumbers as the amplitude of the incoming wave ($B_W$) increases. Therefore, the in-phase branch extends to higher frequencies and the onset of the damping-induced gap in the wavenumber occurs at shorter wavelengths. The onset of the out-of-phase branch, on the other hand, is changed by nonlinearity to a smaller extent, as evident by the imaginary part of the dispersion curve. The overall effect is the gradual decrease in the width of the band gap as the amplitude of the incoming wave increases, until the bandgap is completely closed around $B_W=0.01$.

\begin{figure*}[h!]
    \includegraphics[width=\linewidth]{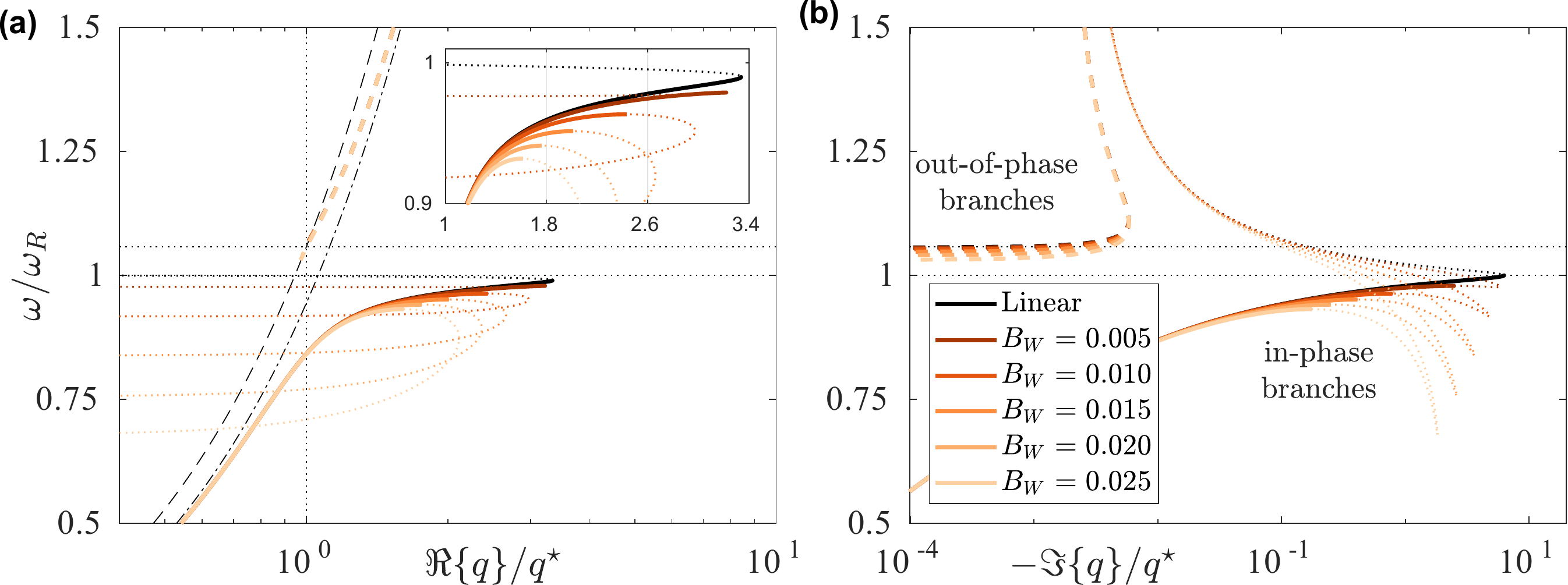}
\caption{Influence of softening, cubic nonlinearity ($\beta=-1$, $k_2=0$) on the dispersion of damped Rayleigh waves with $\zeta=0.010$ for different values of the incoming wave amplitude, $B_W$. Panels (a) and (b) show, respectively, the real and imaginary parts of the dispersion curves. The linear dispersion curve from Fig.~\ref{f:disp_zett} is included for comparison. The portions of the in-phase branches beyond the first turning point are depicted in dotted lines. 
}
\label{f:disp_zett_soft}       
\end{figure*}

Fig.~\ref{f:disp_zett_soft} shows the influence of softening cubic nonlinearity on the dispersion of damped surface waves with $\zeta=0.010$. As already discussed in Sec.~\ref{sec:cubic_soft} and Sec.~\ref{sec:damping_linear}, both damping and softening nonlinearity can lead to the emergence of gaps in admissible wavenumbers in the in-phase dispersion branch. Consistently, we see in Fig.~\ref{f:disp_zett_soft} that increasing the amplitude of the incoming waves decreases the maximum admissible wavenumber. For a linear metasurface with damping, the maximum admissible wavenumber (i.e., the onset of the wavenumber gap) corresponds to the vertical tangency of the dispersion curve ($\partial q/\partial\omega=0$). In the presence of a softening nonlinearity, however, the maximum admissible wavenumber corresponds to the horizontal tangency of the dispersion curve ($\partial \omega / \partial q=0$). This is because the region of the dispersion curve between the horizontal and vertical tangencies corresponds to unstable motion of the local resonators, as already described in Sec.~\ref{sec:cubic_soft}. 
In comparison to the in-phase branches, the out-of-phase branches remain relatively unchanged for weak to moderate values of the incoming wave amplitude; see Fig.~\ref{f:disp_zett_soft}(b). The onset of the out-of-phase branch could eventually move to lower frequencies for large values of the incoming wave amplitude, as already described in Sec.~\ref{sec:cubic_soft} and Fig.~\ref{f:disp_hard}b.


\section{Numerical modeling}
\label{sec:numerics}

\begin{figure}
    \includegraphics[width=\linewidth]{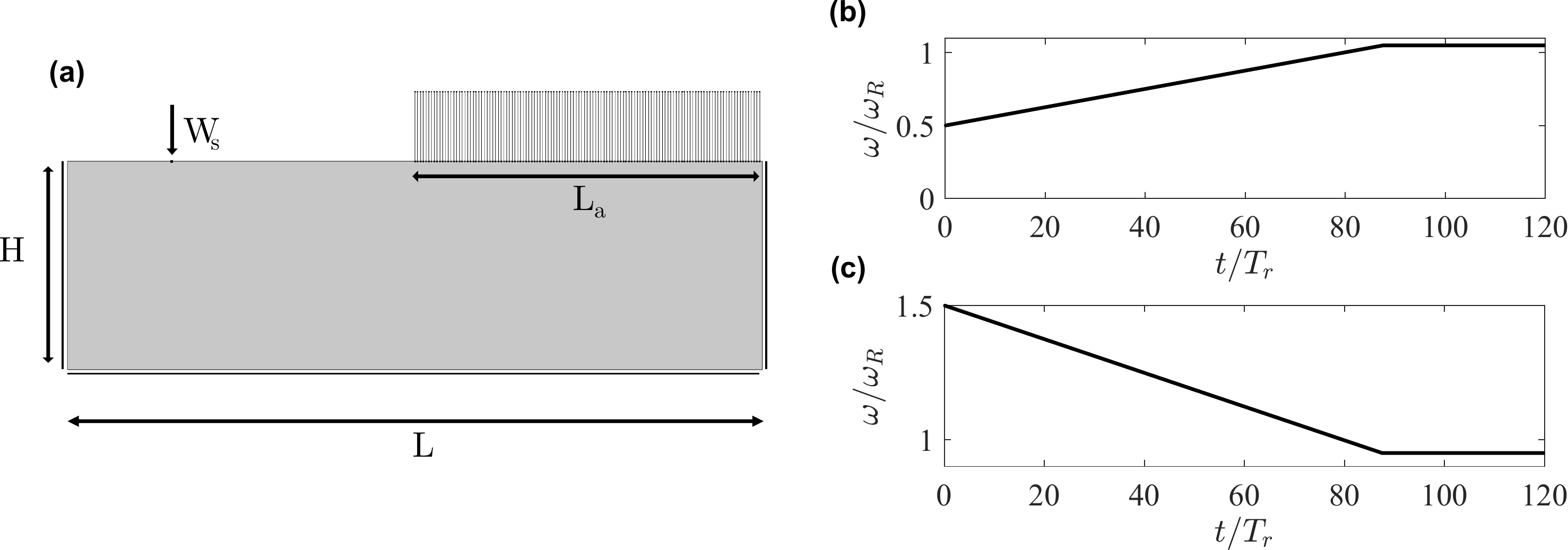}
\caption{FE Numerical model. (a) Geometry of the FE model. (b) Up-chirp  and (c) down-chirp frequency vs. time behavior.$T_R=\frac{2\pi}{\omega_R}$, $t_1=87.5T_R$.}
\label{f:numerical model}       
\end{figure}

We develop a 2D plane-strain finite-element (FE) model comprising an elastic substrate and a finite chain of local resonators at the surface of the substrate (Fig.~\ref{f:numerical model}a). 
The model includes a portion of an elastic substrate of unitary thickness and in-plane dimensions $L\times H$, where  $L=10\lambda_R$, $H=3\lambda_R$, and $\lambda_R=2\pi c_T/\omega_R$. The free surface hosts an array of 125 oscillators,  arranged with a subwavelenght spacing $s=\frac{\lambda_R}{25}$ to cover a total length of $L_a=5\lambda_R$. To replicate the results discussed in the analytical section, we adopt the same mechanical parameters, namely  $c_L/c_T=1.5$ and $\frac{m\omega_r}{\rho Ac_T}=0.15$. 

The wave field is generated by an imposed vertical displacement $W_s(t)$ applied at sufficient distance ($d_s=3.5\lambda_R$) from the first resonator to produce a base excitation mainly conveyed by Rayleigh  modes. Low-reflective boundary conditions (LRB) are used to minimize the amplitude of the reflected signals and simulate an effective semi-infinite domain. 

The FE models are developed in COMSOL Multiphysics using a convergent mesh of linear triangular elements for the substrate and nonlinear truss elements for the resonators. The time-domain integration of the governing equations is performed using the Generalized-$\alpha$ method, employing the damped Newton method as the nonlinear solver at each time step. From the full-field simulations, we collect the displacement response at the base of each resonator $B_W(t,x_r)$ and use them to numerically reconstruct the Rayleigh wave dispersion curves $|B_W(\omega,q)|$ via 2D Fourier transform (2DFFT).

We model the propagation of Rayleigh waves in three different metasurfaces: linear ($\beta=0$), softening ($\beta=-1$) and hardening  ($\beta=1$) type. For each configuration, we perform two time-domain simulations, employing as the input signal $W_s(t)$ a linear chirp with frequency $\omega(t)=\omega_0+ct$, $t=[0,t_1]$, padded by a constant harmonic signal at $\omega_1=\omega(t_1)$. 
For the first batch of simulations, we adopt a chirp with increasing frequency (up-chirp)  between  $\omega_0=0.5\omega_R$ and $\omega_1=1.05\omega_R$. {\color{black}The up-chirp allows to excite the in-phase collective response of the resonators and in turn the acoustic mode supported by the metasurface. In the second batch, we employ a chirp with  decreasing frequency (down-chirp) between $\omega_0=1.5\omega_R$ and $\omega_1=0.95\omega_R$ to excite the out-of-phase, optical,  branch (see Fig.~\ref{f:numerical model}b,c)}.

Fig.~\ref{f:Dispersion numerics} shows the colormaps of the spectral amplitude $|B_W(\omega,q)|$ for up-chirps propagating along a chain of linear (panel a), hardening (panel b) and softening (panel c) resonators. The response of the linear array (panel a), used for reference and validation purposes,  shows evidence of the expected hybridization phenomenon, namely a clear spectral gap and a flat acoustic branch. The numerical results match the analytical predictions (marked by white lines). 

{\color{black} The dispersive characteristics of Rayleigh waves are significantly altered by the nonlinear nature of the interaction force. For a chain of resonators with hardening nonlinearity, Fig.~\ref{f:Dispersion numerics}(b), the spectral gap disappears as predicted by the analysis of Sec.~\ref{sec:cubic_hard}. We observe in the numerical results that the signal amplitude spreads across a wide range of wavenumbers while approaching and crossing the linear resonance $\omega_R$. This phenomenon can be ascribed to the variation of the base amplitude along the chain of resonators, which affects the dispersion of the Rayleigh waves (see Fig.~\ref{f:spectrograms}a).

\begin{figure*}
    \includegraphics[trim={0 0cm 0 0cm}, clip,width=\linewidth]{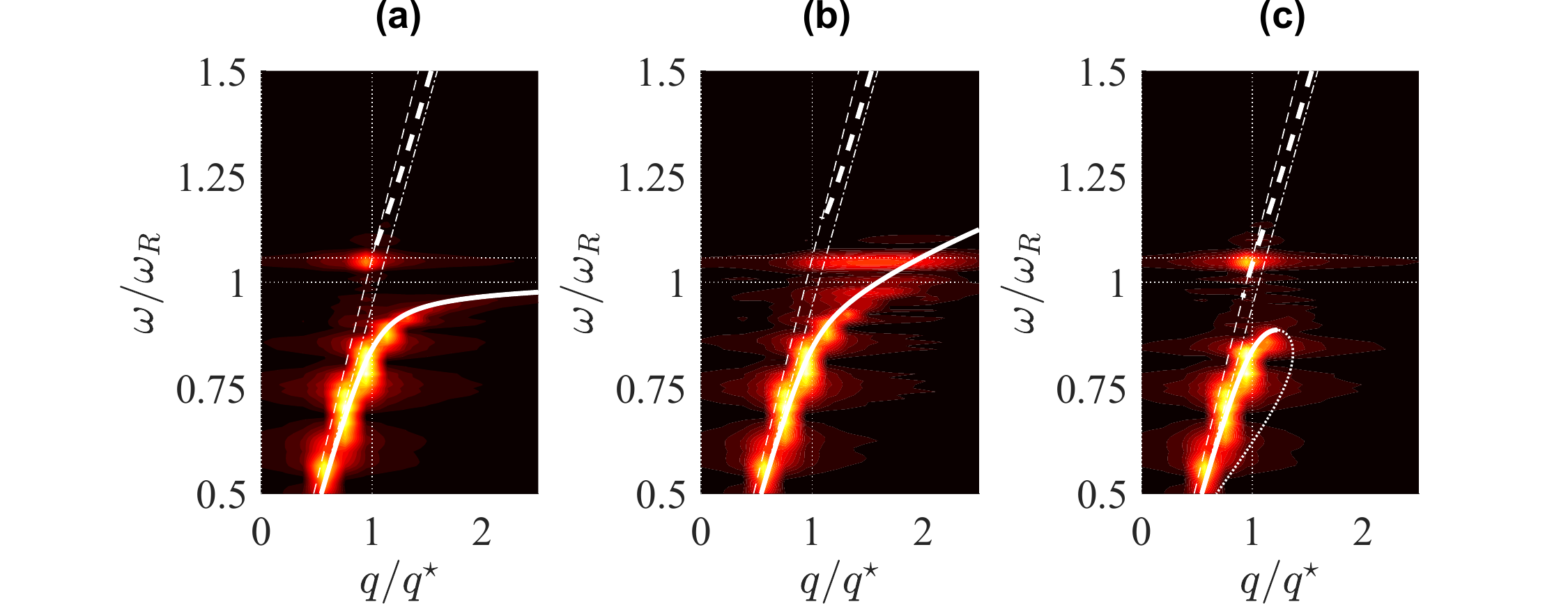}
\caption{Dispersion reconstruction via 2DFFT. Panels (a), (b) and (c) show the reconstructed dispersion curves for the configurations with linear, hardening and softening forces, respectively. The excitation signal is a travelling up-chirp with $\omega=[0.5 - 1.05]\omega_R$. All the models assume a damping coefficient $\zeta=0.010$. The analytical curves, reported as white lines, are obtained assuming a base amplitude displacement $B_W=0.056$.}
\label{f:Dispersion numerics}       
\end{figure*}

In the case of a metasurface with softening nonlinearity, the 2DFFT in Fig.~\ref{f:Dispersion numerics}(c) shows a large spectral gap caused by a shifted and truncated acoustic branch.  As for the hardening configuration, we observe a significant variation in the base amplitude along the metasurface (see Fig.~\ref{f:spectrograms}(b)).  

Nonetheless, for both hardening and softening metasurfaces,  the dispersive features are well captured by the analytical dispersion curves, overimposed to the colormaps in Fig. Fig.~\ref{f:Dispersion numerics}(b),(c), and calculated for a constant reference base displacement amplitude $B_W=0.056$. The value is estimated from the low-frequency base displacement amplitude at the onset of the array. In particular, we consider the time-response at the base of the $10^{th}$ resonator, displayed in Fig.~\ref{f:spectrograms}(c),(d),for hardening and softening metasurfaces, respectively. The choice of this resonator allows to avoid local effects at the interface between the free substrate and the metasurface.

\begin{figure*}
    \includegraphics[trim={0 5.5cm 0 5cm}, clip,width=\linewidth]{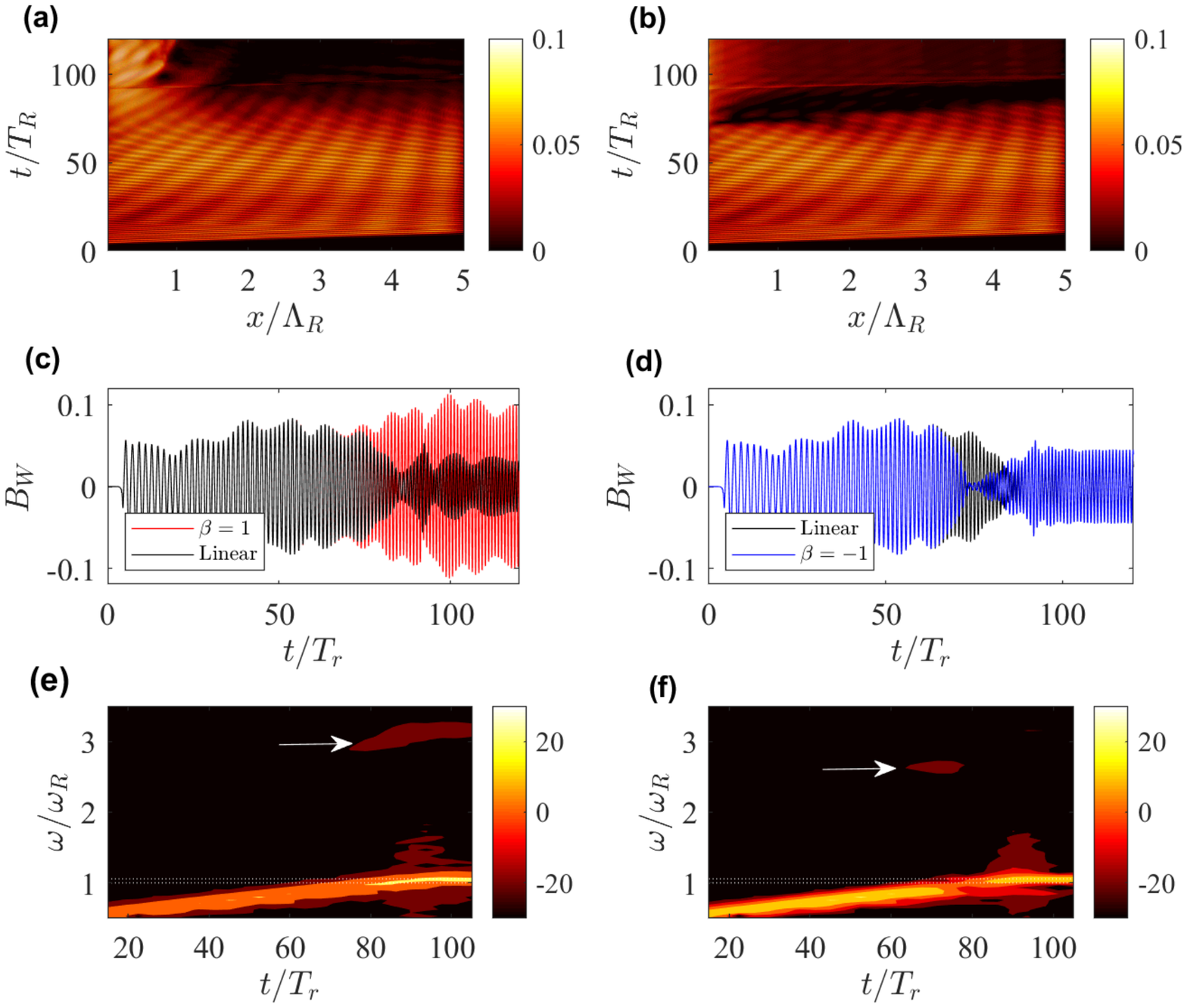}
\caption{Displacement response at the base of the  resonators. Panels (a) and (b) show the base displacement along the entire metasurface for  
the configurations with hardening and softening force, respectively ($\Lambda_R=\frac{2\pi c_T}{\omega_R}$). 
Panels (c) and (d) compare the time-domain response of the hardening (red line) and softening (blue line) configurations, respectively, both extracted at the base of the $10^{th}$ resonator. The response of the linear resonator (black line) is provided for reference. Panels (e) and (f) show the spectrograms (in dB) of the time-domain responses of panels (c) and (d), respectively. The arrows indicates the amplitudes of the third harmonic components. The dashed lines enclose the frequency region of the linear band gap.}
\label{f:spectrograms}       
\end{figure*}

Inspection of the time-domain response can highlight another difference between the effects of the two types of nonlinearity. The response at the base of the hardening resonator (Figs.~\ref{f:spectrograms}(c)) shows no amplitude reduction as a result of the resonance shift towards higher frequency range. Conversely, at the base of the softening resonator, a marked amplitude reduction is observed in a lower and broader frequency range (Figs.~\ref{f:spectrograms}(d))}. These effects are also noticeable in the corresponding spectrograms (short Fourier transform) of the base displacement response, reported in Figs.~\ref{f:spectrograms}(e),(f).
Furthermore, the spectrograms highlight the presence of third harmonics, which are particularly pronounced in the response of the cubic hardening resonators. 

\begin{figure*}
  \includegraphics[trim={0 0cm 0 0cm}, clip,width=\linewidth]{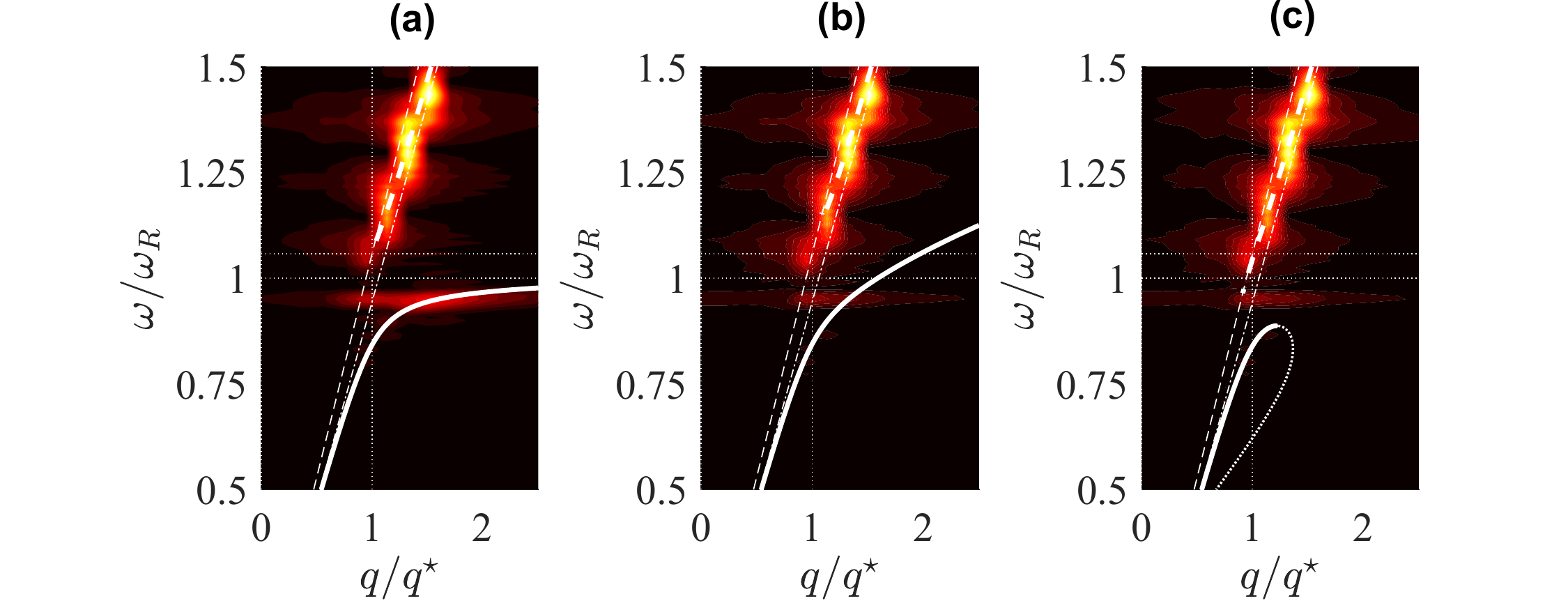}
\caption{Dispersion reconstruction via 2DFFT. Panel (a) reports the reconstructed dispersion curve for a linear chain of oscillators, panels (b) and (c) the dispersion curves for hardening and softening resonators, respectively. The exciting signal is an up-chirp for a travelling  down-chirp with $\omega=[1.5 - 0.95]\omega_R$. All the models assume a damping coefficient $\zeta=0.010$. The analytical curves, reported as white lines, are obtained assuming a base amplitude displacement $B_W=0.056$.}
\label{f:dispersion numeric dec}       
\end{figure*}

Fig.~\ref{f:dispersion numeric dec} shows the reconstructed out-of-phase branch of dispersion curves for metasurfaces with linear (panel a), hardening (panel b), and softening (panel c) behaviour, as obtained from the 2DFFT of the travelling down-chirps. In accordance with the analytical predictions of Sec.~\ref{sec:alldispersion}, the effects of nonlinearity on the out-of-phase branches are weaker than their effects on the in-phase branches. For the range of incoming wave amplitudes investigated here, our numerical analysis cannot clearly capture the weak effects of nonlinearity on the dispersion of out-of-phase branches that were discussed in Sec.~\ref{sec:alldispersion}, partially due to the limited resolution of the 2DFFT.
 
\section{Summary and conclusion}
\label{sec:conclusion}

We provided a theoretical description of the dispersion of Rayleigh waves in a metasurface with nonlinear interaction force between the surface and local resonators. We developed closed-form expressions for the dispersion relation of the metasurface within the context of an amplitude-dependent effective medium. Our analysis uses a single-term plane-wave expansion of the wave field to describe the leading-order effects of nonlinearity in the interaction force, and ignores the generation of harmonics and modal interactions between them. We further investigated the effects of  energy loss in the resonator on transmission of Rayleigh waves. We developed the methodology for a general weak nonlinearity based on a Taylor expansion. Detailed results and discussions, however, were only presented for a interaction force with a symmetric functional form (cubic nonlinearity and linear damping). 

The amplitude-dependent dynamic response of the surface resonators governs the sub-wavelength band gap and manifesting as a shift on the onset or offset of the effective band gap. Specifically, when the metasurface exhibits a hardening type of nonlinearity, the band gap disappears because the in-phase branch of dispersion no longer has an upper bound. Energy loss re-introduces an upper bound to the in-phase dispersion branch. The relative strengths of the nonlinear and damping forces determines if a band gap reappears in this case. The onset of the out-of-phase branch of dispersion shifts to higher frequencies as nonlinearity becomes stronger; this is a marginal effect in comparison to the changes introduced to the in-phase branch. 

When the type of nonlinearity is softening, the in-phase branch terminates at a lower frequency (than the linear scenario), while the out-of-phase branch remains relatively unchanged. Thus, the band gap extends to lower frequencies for moderate strengths of nonlinearity. The onset of the out-of-phase branch shifts to lower frequencies and eventually closes the band gap beyond a certain amplitude of incoming waves. The early termination of the in-phase branch is caused by the dynamic instability of the surface resonators. This is accompanied by the emergence of a maximum admissible wavenumber, thus introducing a spatial gap in the dispersion curve. The maximum admissible wavenumber decreases both with increasing the amplitude of incoming waves and with increasing damping. 

We have used the finite element method to validate our analytical findings and to provide additional insights on the dynamics of the nonlinear metasurface. In particular, we provide evidence of modal interactions in the form of third-harmonic generation and of strong amplitude variation along the metasurface. Beyond validating the predictions of our analytical results, the numerical findings point to promising directions for extension of this work: exploration of the modal interactions by means of higher-order plane-wave expansion, of the spatial gaps in the dispersion relation by reformulating the setup as a scattering problem, and of the additional tuning capabilities introduced by a interaction force with quadratic nonlinearity.

Further analytical developments can facilitate parametric studies of the behavior of the band gap. Experimental investigation of the type of amplitude-dependent behavior described here can pave the way for development of surface wave filtering and control strategies with tunable properties, for example in contact-resonant metasurfaces or in thin-beam metasurfaces with geometric nonlinearity. Beyond that, the present work may contribute to advancements and future developments in seismic metamaterials ~\cite{pastpresentfuture}.

\section*{Conflict of interest}
The authors declare they have no conflict of interest.
\section*{Acknowledgement}
C.D. acknowledges support for this project from NSF EFRI award number 1741565.
A.M. and A.P. acknowledge support for this project from European Union's Horizon 2020 research and innovation programme under the Marie Skłodowska Curie grant agreement No 813424.

\bibliographystyle{spphys}
\bibliography{cicilib.bib}

\begin{thebibliography}{10}
\expandafter\ifx\csname url\endcsname\relax
  \def\url#1{\texttt{#1}}\fi
\expandafter\ifx\csname urlprefix\endcsname\relax\def\urlprefix{URL }\fi
\expandafter\ifx\csname href\endcsname\relax
  \def\href#1#2{#2} \def\path#1{#1}\fi

\bibitem{hussein2014dynamics}
M.~I. Hussein, M.~J. Leamy, M.~Ruzzene, Dynamics of phononic materials and
  structures: Historical origins, recent progress, and future outlook, Applied
  Mechanics Reviews 66~(4) (2014).

\bibitem{ma2016acoustic}
G.~Ma, P.~Sheng, Acoustic metamaterials: From local resonances to broad
  horizons, Science Advances 2~(2) (2016) e1501595.

\bibitem{assouar}
B.~Assouar, B.~Liang, Y.~Wu, Y.~Li, J.-C. Cheng, Y.~Jing, Acoustic
  metasurfaces, Nature Reviews Materials 3~(12) (2018) 460--472.

\bibitem{colombi2016seismic}
A.~Colombi, D.~Colquitt, P.~Roux, S.~Guenneau, R.~V. Craster, A seismic
  metamaterial: The resonant metawedge, Scientific Reports 6~(1) (2016) 1--6.

\bibitem{chaplain2020tailored}
G.~J. Chaplain, J.~M. De~Ponti, A.~Colombi, R.~Fuentes-Dominguez, P.~Dryburg,
  D.~Pieris, R.~J. Smith, A.~Clare, M.~Clark, R.~V. Craster, Tailored elastic
  surface to body wave umklapp conversion, Nature Communications 11~(1) (2020)
  1--6.

\bibitem{dePonti2020}
J.~M. De~Ponti, A.~Colombi, R.~Ardito, F.~Braghin, A.~Corigliano, R.~V.
  Craster, Graded elastic metasurface for enhanced energy harvesting, New
  Journal of Physics 22~(1) (2020) 013013.

\bibitem{addouche2014subwavelength}
M.~Addouche, M.~A. Al-Lethawe, A.~Elayouch, A.~Khelif, Subwavelength
  waveguiding of surface phonons in pillars-based phononic crystal, AIP
  Advances 4~(12) (2014) 124303.

\bibitem{palermo2018control}
A.~Palermo, A.~Marzani, Control of love waves by resonant metasurfaces,
  Scientific Reports 8~(1) (2018) 1--8.

\bibitem{fuentes2021design}
R.~Fuentes-Dom{\'\i}nguez, M.~Yao, A.~Colombi, P.~Dryburgh, D.~Pieris,
  A.~Jackson-Crisp, D.~Colquitt, A.~Clare, R.~J. Smith, M.~Clark, Design of a
  resonant luneburg lens for surface acoustic waves, Ultrasonics 111 (2021)
  106306.

\bibitem{benchabane2019elastic}
S.~Benchabane, A.~Reinhardt, Elastic metamaterials for radiofrequency
  applications, Fundamentals and Applications of Acoustic Metamaterials: From
  Seismic to Radio Frequency 1 (2019) 207--262.

\bibitem{palermo2016engineered}
A.~Palermo, S.~Kr{\"o}del, A.~Marzani, C.~Daraio, Engineered metabarrier as
  shield from seismic surface waves, Scientific Reports 6~(1) (2016) 1--10.

\bibitem{Garova}
E.~Garova, A.~Maradudin, A.~Mayer, Interaction of {R}ayleigh waves with
  randomly distributed oscillators on the surface, Physical Review B 59~(20)
  (1999) 13291.

\bibitem{maznev2015waveguiding}
A.~Maznev, V.~Gusev, Waveguiding by a locally resonant metasurface, Physical
  Review B 92~(11) (2015) 115422.

\bibitem{colquitt2017seismic}
D.~Colquitt, A.~Colombi, R.~Craster, P.~Roux, S.~Guenneau, Seismic
  metasurfaces: Sub-wavelength resonators and rayleigh wave interaction,
  Journal of the Mechanics and Physics of Solids 99 (2017) 379--393.

\bibitem{wootton2019asymptotic}
P.~Wootton, J.~Kaplunov, D.~Colquitt, An asymptotic hyperbolic--elliptic model
  for flexural-seismic metasurfaces, Proceedings of the Royal Society A
  475~(2227) (2019) 20190079.

\bibitem{marigo2020effective}
J.-J. Marigo, K.~Pham, A.~Maurel, S.~Guenneau, Effective model for elastic
  waves propagating in a substrate supporting a dense array of plates/beams
  with flexural resonances, Journal of the Mechanics and Physics of Solids 143
  (2020) 104029.

\bibitem{pu2020seismic}
X.~Pu, A.~Palermo, Z.~Cheng, Z.~Shi, A.~Marzani, Seismic metasurfaces on porous
  layered media: Surface resonators and fluid-solid interaction effects on the
  propagation of rayleigh waves, International Journal of Engineering Science
  154 (2020) 103347.

\bibitem{wu2021non}
Q.~Wu, H.~Chen, H.~Nassar, G.~Huang, Non-reciprocal {R}ayleigh wave propagation
  in space-time modulated surface, Journal of the Mechanics and Physics of
  Solids 146  104196.

\bibitem{palermo2020surface}
A.~Palermo, P.~Celli, B.~Yousefzadeh, C.~Daraio, A.~Marzani, Surface wave
  non-reciprocity via time-modulated metamaterials, Journal of the Mechanics
  and Physics of Solids 145 (2020) 104181.

\bibitem{metawedge}
A.~Colombi, D.~Colquitt, P.~Roux, S.~Guenneau, R.~V. Craster, A seismic
  metamaterial: The resonant metawedge, Scientific Reports 6~(1) (2016) 1--6.

\bibitem{pu2021lamb}
X.~Pu, A.~Palermo, A.~Marzani, Lamb's problem for a half-space coupled to a
  generic distribution of oscillators at the surface, arXiv preprint
  arXiv:2101.09997 (2021).

\bibitem{TournatBertoldi}
X.~Guo, V.~E. Gusev, K.~Bertoldi, V.~Tournat, Manipulating acoustic wave
  reflection by a nonlinear elastic metasurface, Journal of Applied Physics
  123~(12) (2018) 124901.

\bibitem{PalermoCelli}
A.~Palermo, Y.~Wang, P.~Celli, C.~Daraio, Tuning of surface-acoustic-wave
  dispersion via magnetically modulated contact resonances, Physical Review
  Applied 11~(4) (2019) 044057.

\bibitem{Pai}
P.~Wang, J.~Shim, K.~Bertoldi, Effects of geometric and material nonlinearities
  on tunable band gaps and low-frequency directionality of phononic crystals,
  Physical Review B 88~(1) (2013) 014304.

\bibitem{NRM}
H.~Nassar, B.~Yousefzadeh, R.~Fleury, M.~Ruzzene, A.~Al{\`u}, C.~Daraio, A.~N.
  Norris, G.~Huang, M.~R. Haberman, Nonreciprocity in acoustic and elastic
  materials, Nature Reviews Materials 5~(9) (2020) 667--685.

\bibitem{KochmannBertoldi}
D.~M. Kochmann, K.~Bertoldi, Exploiting microstructural instabilities in solids
  and structures: from metamaterials to structural transitions, Applied
  Mechanics Reviews 69~(5) (2017).

\bibitem{bahram}
Y.~Jin, Y.~Pennec, B.~Bonello, H.~Honarvar, L.~Dobrzynski, B.~Djafari-Rouhani,
  M.~Hussein, Physics of surface vibrational resonances: Pillared phononic
  crystals, metamaterials, and metasurfaces, Reports on Progress in Physics
  (2021).

\bibitem{mu2020review}
D.~Mu, H.~Shu, L.~Zhao, S.~An, A review of research on seismic metamaterials,
  Advanced Engineering Materials 22~(4) (2020) 1901148.

\bibitem{WallenNL}
S.~P. Wallen, J.~Lee, D.~Mei, C.~Chong, P.~G. Kevrekidis, N.~Boechler, Discrete
  breathers in a mass-in-mass chain with hertzian local resonators, Physical
  Review E 95 (2017) 022904.

\bibitem{casalotti2018metamaterial}
A.~Casalotti, S.~El-Borgi, W.~Lacarbonara, Metamaterial beam with embedded
  nonlinear vibration absorbers, International Journal of Non-Linear Mechanics
  98 (2018) 32--42.

\bibitem{beamHarmonicGeneration}
X.~Fang, J.~Wen, D.~Yu, G.~Huang, J.~Yin, Wave propagation in a nonlinear
  acoustic metamaterial beam considering third harmonic generation, New Journal
  of Physics 20~(12) (2018) 123028.

\bibitem{lou2020revealing}
J.~Lou, L.~He, J.~Du, H.~Wu, Revealing the linear and nonlinear dynamic
  behaviors of metabeams with a dynamic homogenization model, Journal of
  Vibration and Acoustics 142~(3) (2020).

\bibitem{onemore}
Y.~Xia, M.~Ruzzene, A.~Erturk, Bistable attachments for wideband nonlinear
  vibration attenuation in a metamaterial beam, Nonlinear Dynamics 102~(3)
  (2020) 1285--1296.

\bibitem{graff}
K.~F. Graff, Wave Motion in Elastic Solids, Oxford University Press, 1975.

\bibitem{ewing}
W.~Ewing, W.~Jardetzky, F.~Press, Elastic Waves in Layered Media, McGraw-Hill,
  1957.

\bibitem{Boechler}
N.~Boechler, J.~Eliason, A.~Kumar, A.~Maznev, K.~Nelson, N.~Fang, Interaction
  of a contact resonance of microspheres with surface acoustic waves, Physical
  Review Letters 111~(3) (2013) 036103.

\bibitem{NayfehMook}
A.~Nayfeh, D.~Mook, Nonlinear Oscillations, John Wiley \& Sons, 1979.

\bibitem{PalermoSeb}
A.~Palermo, S.~Kr{\"o}del, A.~Marzani, C.~Daraio, Engineered metabarrier as
  shield from seismic surface waves, Scientific Reports 6~(1) (2016) 1--10.

\bibitem{WallenMaznev}
S.~P. Wallen, A.~A. Maznev, N.~Boechler, Dynamics of a monolayer of
  microspheres on an elastic substrate, Physical Review B 92~(17) (2015)
  174303.

\bibitem{Djafar2021}
R.~Cai, Y.~Jin, T.~Rabczuk, X.~Zhuang, B.~Djafari-Rouhani, Propagation and
  attenuation of {R}ayleigh and pseudo surface waves in viscoelastic
  metamaterials, Journal of Applied Physics 129~(12) (2021) 124903.

\bibitem{GeneralMead}
D.~Mead, A general theory of harmonic wave propagation in linear periodic
  systems with multiple coupling, Journal of Sound and Vibration 27~(2) (1973)
  235--260.

\bibitem{mikeJAP}
M.~I. Hussein, M.~J. Frazier, Band structure of phononic crystals with general
  damping, Journal of Applied Physics 108~(9) (2010) 093506.

\bibitem{leamy}
F.~Farzbod, M.~J. Leamy, Analysis of {B}loch’s method in structures with
  energy dissipation, Journal of Vibration and Acoustics 133~(5) (2011).

\bibitem{laudePRB1}
R.~P. Moiseyenko, V.~Laude, Material loss influence on the complex band
  structure and group velocity in phononic crystals, Physical Review B 83~(6)
  (2011) 064301.

\bibitem{jensenJVA}
E.~Andreassen, J.~S. Jensen, Analysis of phononic bandgap structures with
  dissipation, Journal of Vibration and Acoustics 135~(4) (2013).

\bibitem{Maznev}
A.~Maznev, Bifurcation of avoided crossing at an exceptional point in the
  dispersion of sound and light in locally resonant media, Journal of Applied
  Physics 123~(9) (2018) 091715.

\bibitem{FangDamped}
N.~Fang, D.~Xi, J.~Xu, M.~Ambati, W.~Srituravanich, C.~Sun, X.~Zhang,
  Ultrasonic metamaterials with negative modulus, Nature Materials 5~(6) (2006)
  452--456.

\bibitem{pastpresentfuture}
S.~Br{\^u}l{\'e}, S.~Guenneau, Past, present and future of seismic
  metamaterials: experiments on soil dynamics, cloaking, large scale analogue
  computer and space--time modulations, Comptes Rendus. Physique 21~(7-8)
  (2020) 767--785.

\end{thebibliography}

\end{document}